\documentclass[utf8]{frontiersFPHY} 

\setcitestyle{square} 
\usepackage{url,hyperref,microtype,subcaption}
\usepackage[onehalfspacing]{setspace}
\usepackage[displaymath]{lineno}
\usepackage{journalabb}

\makeatletter
\let\LN@gather\gather
\let\LN@endgather\endgather
\renewcommand{\gather}{\linenomath\LN@gather}
\renewcommand{\endgather}{\LN@endgather\endlinenomath}
\makeatother

\def\keyFont{\fontsize{8}{11}\helveticabold }
\def\firstAuthorLast{Nakanotani {et~al.}} 
\def\Authors{Masaru Nakanotani\,$^{1,*}$, Gary P. Zank\,$^{1,2}$ and Lingling Zhao\,$^{1,2}$}
\def\Address{$^{1}$Center for Space Plasma and Aeronomic  Research (CSPAR), University of Alabama in Huntsville, Huntsville, AL, USA \\
	$^{2}$Department of Space Science, University of Alabama in Huntsville, Huntsville, AL, USA  }
\def\corrAuthor{Masaru Nakanotani}

\def\corrEmail{mn0052@uah.edu}

\begin{document}
	\onecolumn
	\firstpage{1}
	
	\title[Multiple current sheets and particle acceleration]{Particle acceleration in an MHD-scale system of multiple current sheets}
	
	\author[\firstAuthorLast ]{\Authors} 
	\address{} 
	\correspondance{} 
	
	\extraAuth{}

	\maketitle

	\begin{abstract}
		\section{}
		We investigate particle acceleration in an MHD-scale system of multiple current sheets by performing 2D and 3D MHD simulations combined with a test particle simulation.
		The system is unstable for the tearing-mode instability, and magnetic islands are produced by magnetic reconnection.
		Due to the interaction of magnetic islands, the system turns into a turbulent state.
		The 2D (3D) case yields both $-5/3$ ($-11/3$ and $-7/3$) power-law spacetra for magnetic and velocity fluctuations.
		Particles are efficiently energized by the generated turbulence, and it forms a power-law tail with an index of $-2.2$ and $-4.2$ in the energy distribution function for the 2D and 3D case, respectively.
		We find more energetic particles outside magnetic islands rather than inside.
		We observe super-diffusion in the 2D ($\sim t^{2.27}$) and 3D  ($\sim t^{1.2}$) case in the energy space of energetic particles.
		

		\tiny
		\keyFont{ \section{Keywords:} MHD simulation, Multiple current sheets, Turbulence, Particle acceleration} 
	\end{abstract}
	
	\section{Introduction}
	Magnetic reconnection at a current sheet is a fundamental process in plasma physics \cite{biskamp1994,yamada2010,hesse2020}.
	Magnetic reconnection can be characterized as a topological change of anti-parallel magnetic fields where the frozen-in condition is broken.
	The reconnected magnetic field drags plasma away due to the tension force.
	The outflow speed roughly corresponds to the Alfv\'en speed. 
	As a result of magnetic reconnection, two separated plasmas are mixed together.
	
	It is believed that magnetic reconnection is capable of generating energetic particles \cite{blandford2017}.
	Several mechanisms have been proposed so far: (1) Speiser (meandering) motion across anti-parallel magnetic fields directly accelerates particles by the inductive electric field \cite{speiser1965}, (2) particles gain energy due to the conservation of the first adiabatic at the pileup region of magnetic field \cite{hoshino2001}, (3) Fermi-type acceleration occurs due to the contraction of the magnetic islands \cite{drake2006,oka2010,xiaocan2021}. 
	Several kinetic simulations show the existence of non-thermal particles forming a power-law tail in the energy distribution function associated with the evolution of magnetic reconnection \cite{sironi2014,guo2014,dahlin2014,werner2015,li2017,zhang2021,arnold2021}.
	
	The system of multiple current sheets has been paid attention to in recent years.
	There are several places forming multiple current sheets in the heliosphere.
	For instance, heliospheric current sheets (HCSs) \cite{smith2001} are usually stable in the solar wind, but they are compressed at the heliospheric termination shock and can be unstable in the heliosheath.
	Spacecraft observations across the sector boundaries often find multiple thin current sheets inside a HCS, and these can be interpreted as the folding of individual magnetic flux tubes \cite{crooker1993,dahlburg1995,maiewski2020}.
	Not only in the heliosphere, but pulsar winds also have a similar structure, and it is believed that the interaction of the current sheets with the pulsar termination shock produces energetic particles and is responsible for conversion of Poynting dominated outflows to the observed radiation via energetic particles produced by the interaction \cite{lyubarsky2005,nagata2008,sironi2011,cerutti2020,lu2021}.
	
	Once those current sheets become unstable, it is thought that the system produces several magnetic islands due to magnetic reconnection and then turns into a turbulent state.
	\citet{zhang2011,akramov2017} performed MHD simulations of double current sheets and showed that growing magnetic islands interact with each other and then the system appears to be turbulence.
	\citet{gengell2015,burgess2016} performed 3D hybrid kinetic simulations of multiple current sheets.
	The system is unstable for the tearing-mode and drift-kink instability, these instabilities lead the system to a turbulent state with a power-law spectrum with an index of $-7/3$ for magnetic fluctuations. 
	
	Particle acceleration among multiple magnetic islands has been proposed as an efficient acceleration process.
	\citet{zank2014,zank2015,leroux2015} model a focused Parker transport equation including the effects of electric field induced by magnetic island reconnection and magnetic island contraction to understand the flux of anomalous cosmic rays observed by Voyager spacecraft, which continuously increases in the downstream of the heliospheric termination shock.
	The model successfully reproduces the observed flux and shows that the energy spectrum becomes harder because of acceleration by magnetic islands in the downstream of a shock wave \cite{zank2015,zhao2019}.
	This has been also observed at interplanetary shock waves at $5$ au \cite{zhao2018,adhikari2019}.
	
	However, recent kinetic simulations of multiple current sheets for a non-relativistic plasma did not show very efficient particle acceleration as expected by models.
	\citet{drake2010} performed 2D full PIC simulations of multiple current sheets and observed particle energization by a few decades in energy, but a power-law energy distribution did not form.
	3D hybrid kinetic simulations was done by \citet{burgess2016}, and apparent particle acceleration of ions and pickup ions was not found.
	\citet{nakanotani2021} investigated the interaction of current sheets with a shock wave and found ion flux increase associated with the evolution of the tearing-mode instability of current sheets in the downstream of the shock wave.
	However, the power-law index of the energy spectrum unchanged associated with the generation of multiple islands due to the tearing-mode instability.
	Note that particle acceleration in multiple current sheets of a relativistic electron-positron plasma has been shown efficient \cite{hoshino2012}.
	
	One question arises: how efficient is particle acceleration on a larger scale, say, MHD scale?
	Recently, \citet{arnold2021} show that electrons are efficiently accelerated by Fermi acceleration due to the coalescence of magnetic islands using MHD simulations combined with guiding-center approximated electrons and including kinetic effects of energetic electrons.
	They pointed out that standard PIC simulations yield only a short power-law tail which extends a decade in energy because of the limitation of the simulation size.
	Therefore, this can be a reason why particle acceleration in the previous studies of multiple current sheets is not efficient as expected.
	We pursue to answer if particle acceleration on a larger scale is efficient or not.
	
	
	In this study, we combine MHD simulations and test particle simulations to investigate particle acceleration.
	This method has been used for several investigations of particle acceleration in magnetic reconnection and turbulence for non-relativistic \cite{matthaeus1984,ambrosiano1988,dmitruk2003,dmitruk2004} and relativistic particles \cite{kowal2012,pezzi2022}.
	Although they ignore a feedback effect from energetic particles to MHD simulations, they provide valuable insights for particle acceleration in the MHD scale, which can not be easily obtained in kinetic simulations due to computational limitation.
	A similar idea has been applied for test-particle electrons in hybrid kinetic simulations \cite{guo2010,trotta2020}.
	We perform 2D and 3D MHD simulations of multiple current sheets combined with test particle simulations.
	
	This paper is organized as follows.
	In Section 2, we describe the scheme of an MHD simulation combined with a test particle simulation and initial conditions.
	Section 3 shows results of 2D and 3D simulations that present the evolution of multiple current sheets, particle acceleration, and particle diffusion in the energy space.
	The last section provides the discussion and conclusion to derive that particle acceleration in MHD-scale multiple current sheets is indeed efficient in both 2D and 3D systems.
	
	
	
	
	\section{Method: MHD + Test Particle Simulation}
	We combine an MHD simulation with a test particle simulation to investigate particle acceleration in a system of multiple current sheets.
	We solve the following compressible ideal-MHD equations,
	\begin{gather}
		\partial_t\rho+\nabla\cdot\left(\rho {\bf V}\right)=0; \\
		\partial_t\left(\rho{\bf V}\right)+\nabla\cdot\left(\rho{\bf VV}+P^*-{\bf BB}\right)=0; \\
		\partial_t e+\nabla\cdot\left(h{\bf V}+{\bf E}\times{\bf B}\right)=0; \\
		\partial_t{\bf B}+\nabla\times{\bf E}=0; \\
		P^*=P+\frac{1}{2}\left({\bf B}\cdot{\bf B}\right); \\
		e=\frac{P}{\gamma-1}+\frac{1}{2}\rho\left({\bf V}\cdot{\bf V}\right)+\frac{1}{2}\left({\bf B}\cdot{\bf B}\right); \\
		h=\frac{\gamma}{\gamma-1}P+\frac{1}{2}\rho\left({\bf V}\cdot{\bf V}\right); \\
		{\bf E}=-{\bf V}\times{\bf B}+\eta{\bf J},
	\end{gather}
	where $\rho$ is the plasma density, ${\bf V}$ plasma velocities, $P$ plasma pressure, ${\bf B}$ magnetic fields, ${\bf E}$ electric fields, $\eta$ artifitical magnetic resistivity, and ${\bf J}$ current densities.
	$\gamma$ is an adiabatic index, and we set $\gamma=5/3$.
	
	We use an MHD scheme proposed by \citet{kawai2013}.
	The first spatial derivative is calculated by the sixth-order compact scheme, and time integration is done by the third-order total variation diminishing (TVD) Runge–Kutta scheme \cite{shu1988}.
	The artifitial magnetic resistivity has the following form \cite{kawai2013},
	\begin{equation}
		\eta=C_\eta\overline{\frac{1}{\rho c_s}\left|\sum_{l=1}^3\frac{\partial^4\left|{\bf J}\right|^2}{\partial \chi^4_l}\Delta \chi^4_l\Delta^3\right|},
	\end{equation}
	where $C_\eta$ is an dimensionless and arbitrary parameter, $c_s$ the local sound speed, $\chi_l$ referes to the Cartesian coordinates in the $l$-direction, and $\Delta \chi_l$ is the lobal grid spacing in the $l$-direction.
	Here, we set $\Delta=\sqrt{\Delta x^2+\Delta y^2+\Delta z^2}$, which gives larger amount of magnetic resistivity compared to the form in \cite{kawai2013}.
	The overbar denotes an approximate truncated Gaussian filter \cite{cook2004}.
	We use a fourth-order explicit scheme \cite{kawai2008} for the fourth derivative.
	The magnetic resistivity with this form automatically localizes in regions where the current density has a strong gradient, such as current sheets.
	Therefore, the resistivity tends to damp turbulence less than a constant magnetic resistivity.
	We also introduce an artificial bulk viscosity and mass diffusivity to capture a shock wave and contact discontinuity correctly \cite{kawai2013}.
	We note that the divergence-free condition ($\nabla\cdot{\bf B}=0$) is satisfied at around machine accuracy ($\sim10^{-13}$) since we use a central-type finite difference scheme \cite{toth2000,kawai2013}.
	
	We set multiple current sheets in a periodic box.
	We assume the force-free condition for the current sheets \cite{bobrova2001,nishimura2003,du2020},
	\begin{gather}
		B_x=B_0\tanh\left[\frac{d}{\pi L_0}\sin\left(\frac{\pi y}{d}\right)\right]; \\
		B_y=0; \\
		B_z=B_0\sqrt{1+\left(\frac{B_g}{B_0}\right)^2-\left(\frac{B_x}{B_0}\right)^2},
	\end{gather}
	where $B_0$ is the in-plane magnetic field, $d$ is the distant between two neighboring current sheets, $L_0$ the half thickness of a current sheet, and $B_g$ the background magnetic field.
	The plasma density and pressure are set to be uniform.
	We add small fluctuations $\delta A_z$ in the $z-$componet of the vector potential to initiate magnetic reconnection at current sheets for 2D and 3D simulations,
	\begin{gather}
		\delta A_z^{2D}=\sum_{m_x=-5}^5\sum_{m_y=-5}^5\delta A_0\cos\left(m_x\frac{2\pi}{L_x}x+m_y\frac{2\pi}{L_y}y+\phi^{2D}(m_x,m_y)\right); \\
		\delta A_z^{3D}=\sum_{m_x=-5}^5\sum_{m_y=-5}^5\sum_{m_z=-5}^5\delta A_0\cos\left(m_x\frac{2\pi}{L_x}x+m_y\frac{2\pi}{L_y}y+m_z\frac{2\pi}{L_z}+\phi^{3D}(m_x,m_y,m_z)\right), \label{eq:3Dfluct}
	\end{gather} 
	where $\delta A_0$ is a constant value, and $\phi^{2D}$ and $\phi^{3D}$ are random phases for 2D and 3D simulations, respectively.
	We use $\delta A_0=0.05$ and $0.02$ for the 2D and 3D case, respectively.
	We confirmed that the overall evolution of the current sheets was similar as uniform random fluctuations were used and, therefore, it does not depend on the choice of initial fluctuations.
	
	Simulation parameters used in the MHD simulations are as follows.
	We use $L_0$ as the unit length of the simulation and the alfv\'en speed $v_{A0}$ defined by $B=\sqrt{B_0^2+B_g^2}$ as the unit speed so that $L_0=1$ and $v_{A0}=1$. 
	We also set the uniform plasma density to $\rho_0=1$.
	The size of the simulation box is $L_x\times L_y=160L_0\times40L_0$ with the grid number $N_x\times N_y=1024\times256$ and $L_x\times L_y\times L_z=160L_0\times40L_0\times40L_0$ with the grid number $N_x\times N_y\times N_z=1024\times256\times256$ for 2D and 3D simulations, respectively.
	The total plasma beta ($\beta=\beta_i+\beta_e$) corresponds to $1$. 
	Here, $\beta_i$ and $\beta_e$ are the ion and electron plasma beta, respectively.
	We set the parameter $C_\eta=2$ for both 2D and 3D simulations. 
	We put $4$ current sheets in the box ($d=10L_0$).
	The Courant-Friedrichs-Lewy (CFL) number is $0.5$ and $0.25$ for 2D and 3D simulations, respectively.
	In this study, we only consider cases without the background magnetic field ($B_g=0$).
	
	At the same time, we solve the following equation of motion in a normalized form for non-relativistic particles using the standard Buneman-Boris method,
	\begin{equation}
		\frac{d{\bf v}}{dt}=\alpha\left({\bf v}-{\bf V}\right)\times{\bf B}.
	\end{equation}
	Here, $\alpha=T_0\Omega_c$ where $T_0$ is the characteristic time scale of the MHD simulation and $\Omega_c$ is the cyclotron frequency of particles.
	The parameter $\alpha$ is an arbitrary and user-specified parameter since the system of the ideal MHD is scale-free, and we set $\alpha=500$.
	The same normalization used in the MHD simulation is applied in the equation of motion, and the particle energy is normalized by $E_0=m_p v_{A0}^2$ where $m_p$ is the particle mass.
	The total number of particles is $N_p=5*1024*256$ and $1024*256*256$ for 2D and 3D simulations, respectively.
	We distribute particles uniformly in space, and they have the Maxwell distribution for velocities with a temperature of $T_p=0.25$.
	Here, we assume the equal temperature between ions and electrons.
	We introduce sub-cycles when calculating the equation of motion with a time step of $\Delta t_p=\Delta t_{MHD}/250$ where $\Delta t_{MHD}$ is the time step calculated in the MHD simulation since the MHD time step can be larger than the cyclotron period.
	Although we do not have to specify if the test particle simulation in the MHD simulation is for electrons or ions, the parameter $\alpha=500$ can be appropriate for ions rather than electrons since $\alpha$ may become much larger for electrons in the scale of our interest \cite{dmitruk2003}.
	
	\section{Results}
	\subsection{2D case}
\begin{figure}[h!]
		\begin{center}
			\includegraphics[width=16cm]{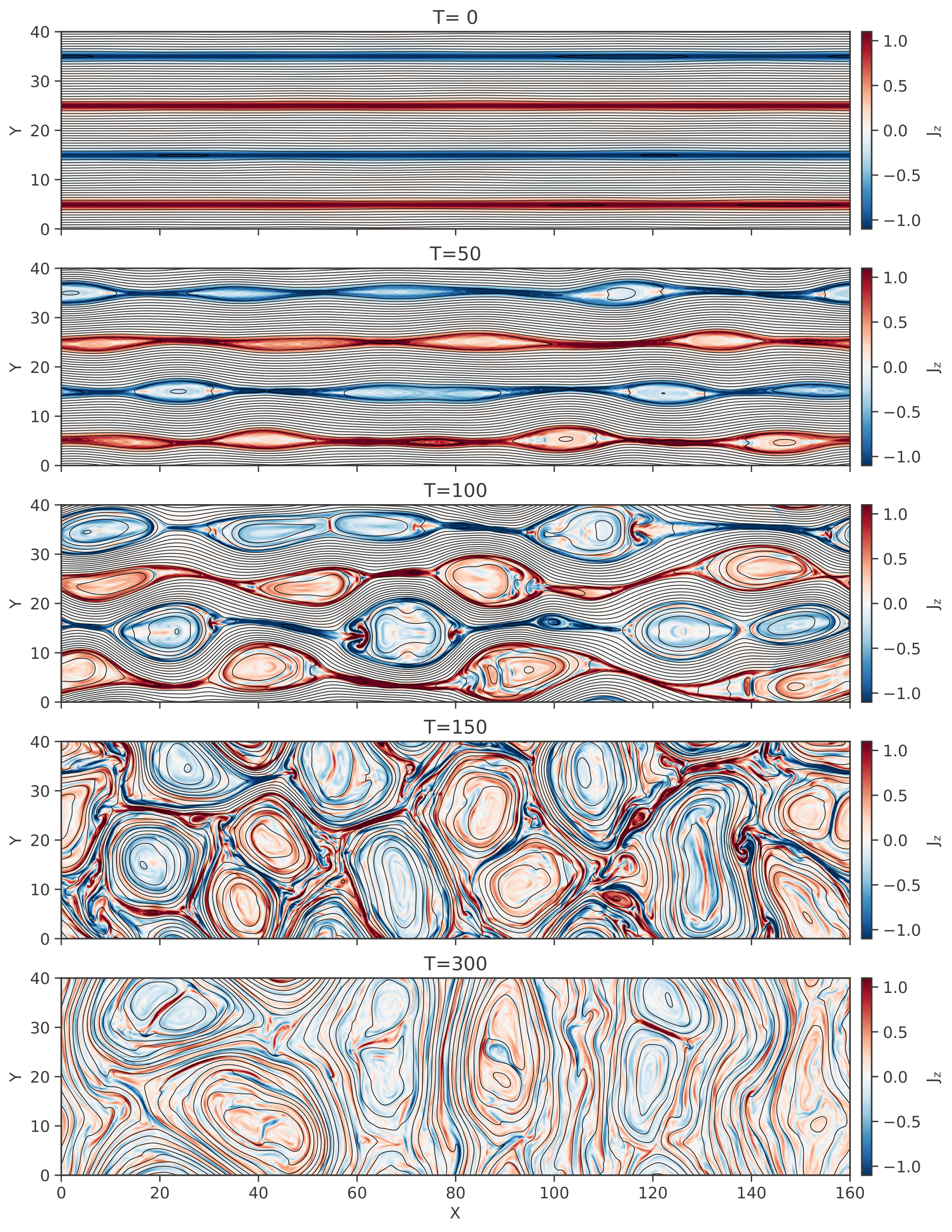}
		\end{center}
		\caption{Snapshots of the current density $J_z$ in the 2D case at different times, $t=0, 50, 100, 150, 300$ from top to bottom. Black lines represent the magnetic field lines (contour lines for the vector potential $A_z$).}
		\label{fig:2Dsnapshot}
	\end{figure}

	Multiple current sheets evolve into a turbulent state.
	Fig. \ref{fig:2Dsnapshot} shows the time evolution of the current density $J_z$ from $t=0$ to $300$.
	The black lines show the magnetic field lines 
	There are $4$ current sheets located equidistant at the initial time.
	The added initial fluctuations onset the tearing-mode instability, and we can see that magnetic reconnection occurs in the current sheets at $t=50$.
	Since the phase of the fluctuations is random, the location of magnetic reconnection is also random.
	As the simulation proceeds, magnetic islands produced by the magnetic reconnection grow in size and merge with each other in the same current sheet. 
	When the size of magnetic islands is roughly equal to or larger than the initial current sheet distance ($10L_0$), magnetic islands start interacting ($t=150$).
	We observe that regions outside the magnetic islands become turbulent.
	At the later time ($t=300$), the size of merging islands becomes around $20L_0$, and the system turns into turbulence.
	
\begin{figure}[h!]
		\begin{center}
			\includegraphics[width=15cm]{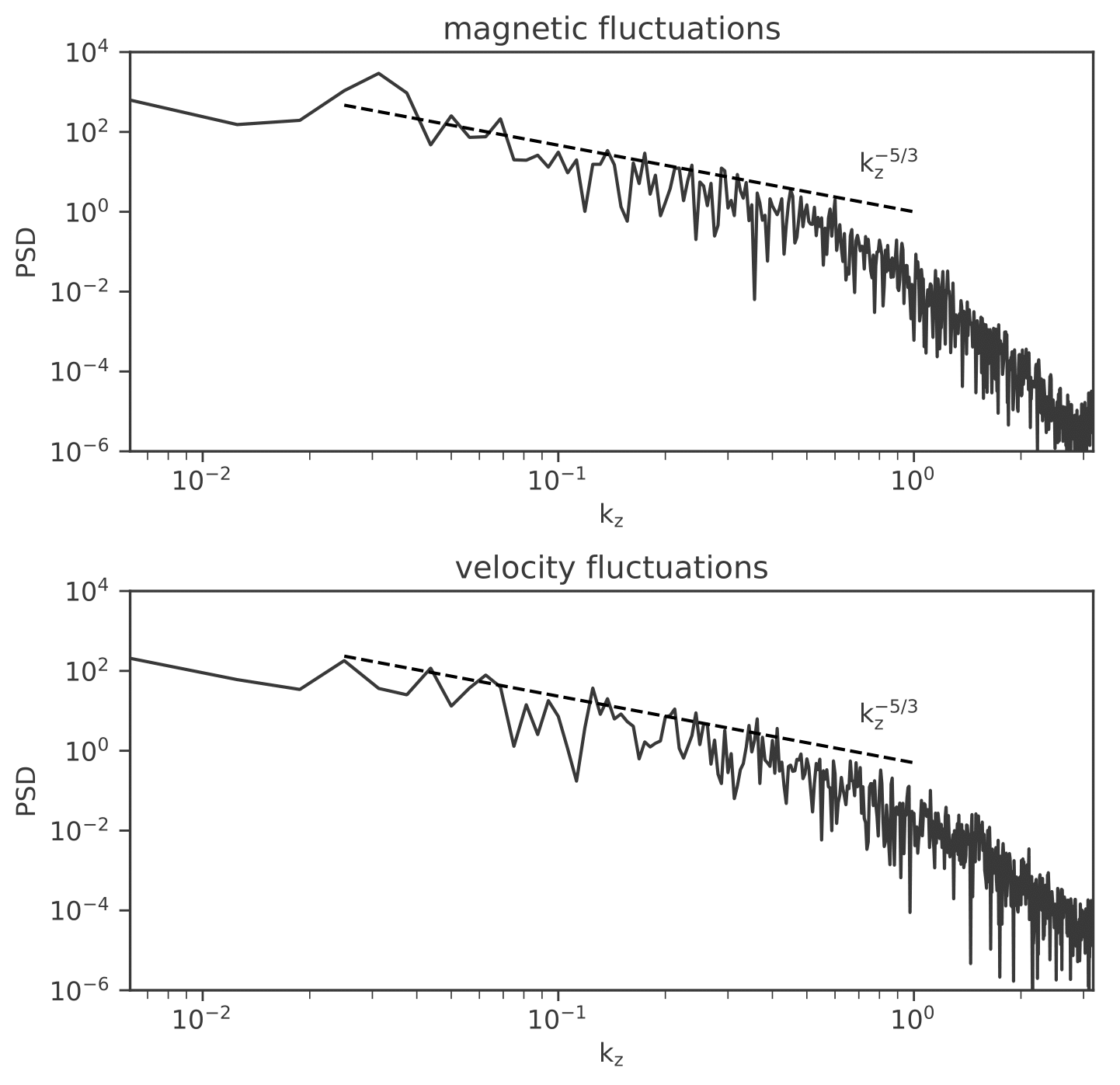}
		\end{center}
		\caption{Power spectrum density of magnetic and velocity fluctuations of the 2D case at $t=300$. }
		\label{fig:2DPSD}
	\end{figure}

	The turbulence exhibits a $-5/3$ power-law in the magnetic and velocity fluctuations.
	Fig. \ref{fig:2DPSD} shows the power spectrum density (PSD) of magnetic ($B=\sqrt{B_x^2+B_y^2+B_z^2}$) and velocity ($V=\sqrt{V_x^2+V_y^2+V_z^2}$) fluctuations in the $z-$direction averaged along the $y-$direction at $t=300$.
	The power-law index of both PSDs can be fitted by $-5/3$ over the range of $k_z\in [0.02, 0.5]$.
	The larger wavenumber region is damped, and this is because of dissipations due to the artificial magnetic resistivity and bulk viscosity included to stabilize the simulation. 
	The normalized cross helicity $<\sigma_c>$ and normalized residual energy $<\sigma_r>$ \cite{zank2012} averaged over the simulation domain at $t=300$ are $0.017$ and $-0.61$, respectively.
	This suggests that the energy of velocity and magnetic fluctuations in forward and backward fluctuations is roughly equal, and the Alfv\'enic fluctuations ($V_A=B/\sqrt{4\pi\rho}$) are stronger than the velocity fluctuations.
	The PSDs also confirm that the later stage of the system is in a turbulent state.
	
\begin{figure}[h!]
		\begin{center}
			\includegraphics[width=15cm]{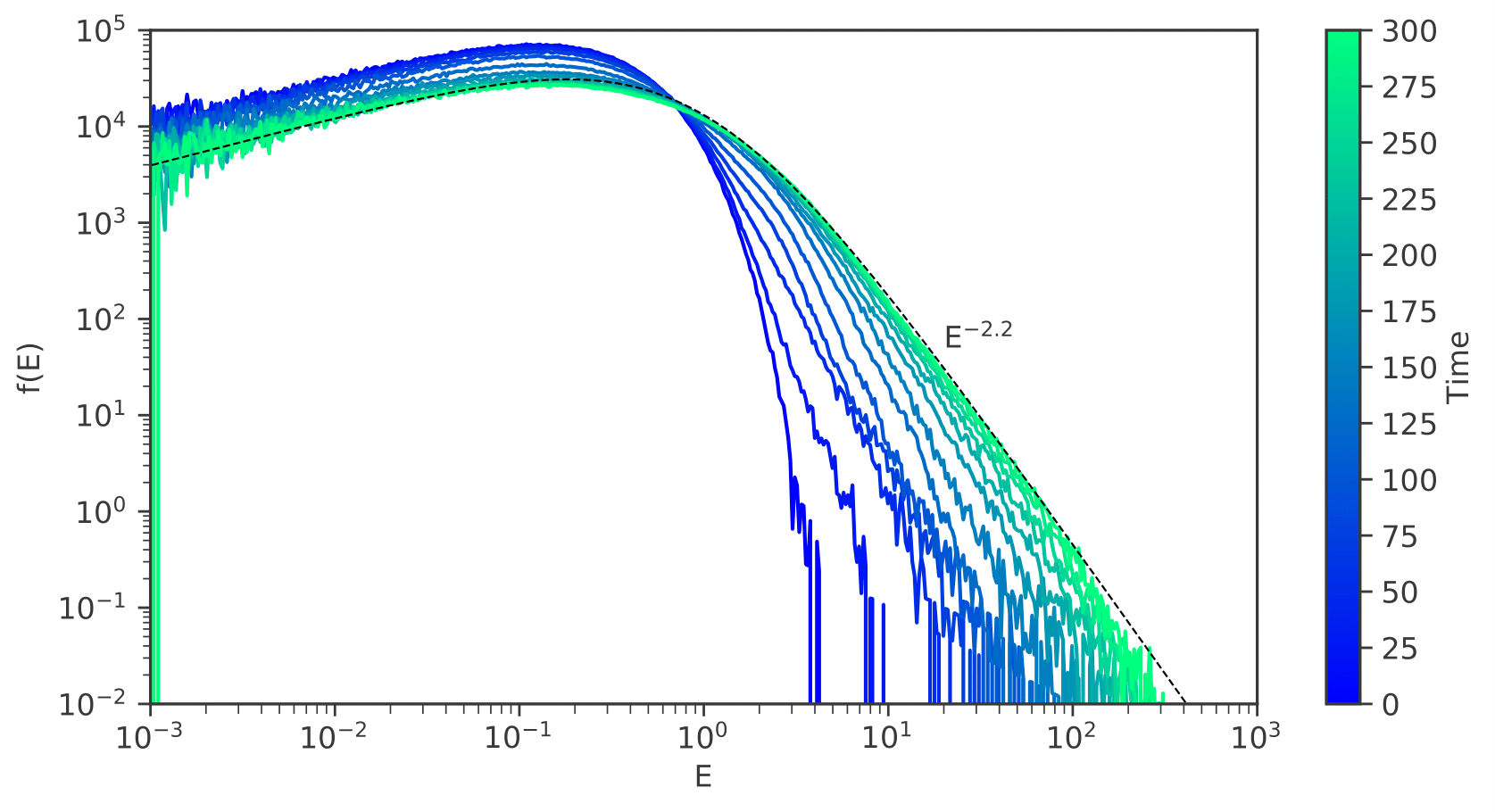}
		\end{center}
		\caption{Time evolution of the energy distribution of test particles in the 2D case. Black dashed line is a Kappa distribution with a temperature of $T_\kappa=1.2$ and a Kappa index of $\kappa=2.2$.}
		\label{fig:2Dedist}
	\end{figure}

	Non-thermal particles are produced during the evolution of the multiple current sheets into turbulence.
	Fig. \ref{fig:2Dedist} shows the time evolution of the energy distribution function of particles.
	We use all particles in the simulation domain to calculate an energy distribution function.
	At $t=0$, the distribution is a Maxwell distribution with a temperature of $T_p=0.25$.
	We can see that a non-thermal tail forms at $t=25$ and $50$. 
	These times correspond to the onset of magnetic reconnection at current sheets.
	In the later times, non-thermal particles are further produced especially after magnetic islands start interacting ($t=150$), and also the distribution is heated.
	At the end of the simulation time ($t=300$), the distribution has a clear non-thermal and power-law tail with an index of $-2.2$.
	The final distribution can be fitted by a Kappa distribution \cite{livadiotis2013},
	\begin{equation}
		f(E)=\frac{2N_\kappa\sqrt{E}}{\sqrt{\pi(k_BT_\kappa)^3}}\frac{\Gamma(\kappa+1)}{(\kappa-3/2)^{3/2}\Gamma(\kappa-1/2)}\left[1+\frac{E}{k_BT_\kappa(\kappa-3/2)}\right]^{-(\kappa+1)}, \label{eq:kappa}
	\end{equation}
	where $N_\kappa$ is the number of particles, $k_B$ the Boltzmann constant, $T_\kappa$ the kappa temperature, $\Gamma$ the Gamma function, $\kappa$ the Kappa (or power-law) index.
	The black dashed line is a Kappa distribution with a temperature of $T_\kappa=1.2$ and $\kappa=2.2$.
	We can clearly see that the power-law tail of the simulated energy distribution at $t=300$ is fitted well by the Kappa distribution over the range of $E\in[1,100]$.
	The maximum energy of accelerated particles is $\sim300E_0$.
	Energetic particles are produced during the evolution from the onset of magnetic reconnection to turbulence, and the final distribution has a power-law tail with an index of $-2.2$.
	
\begin{figure}[h!]
		\begin{center}
			\includegraphics[width=16cm]{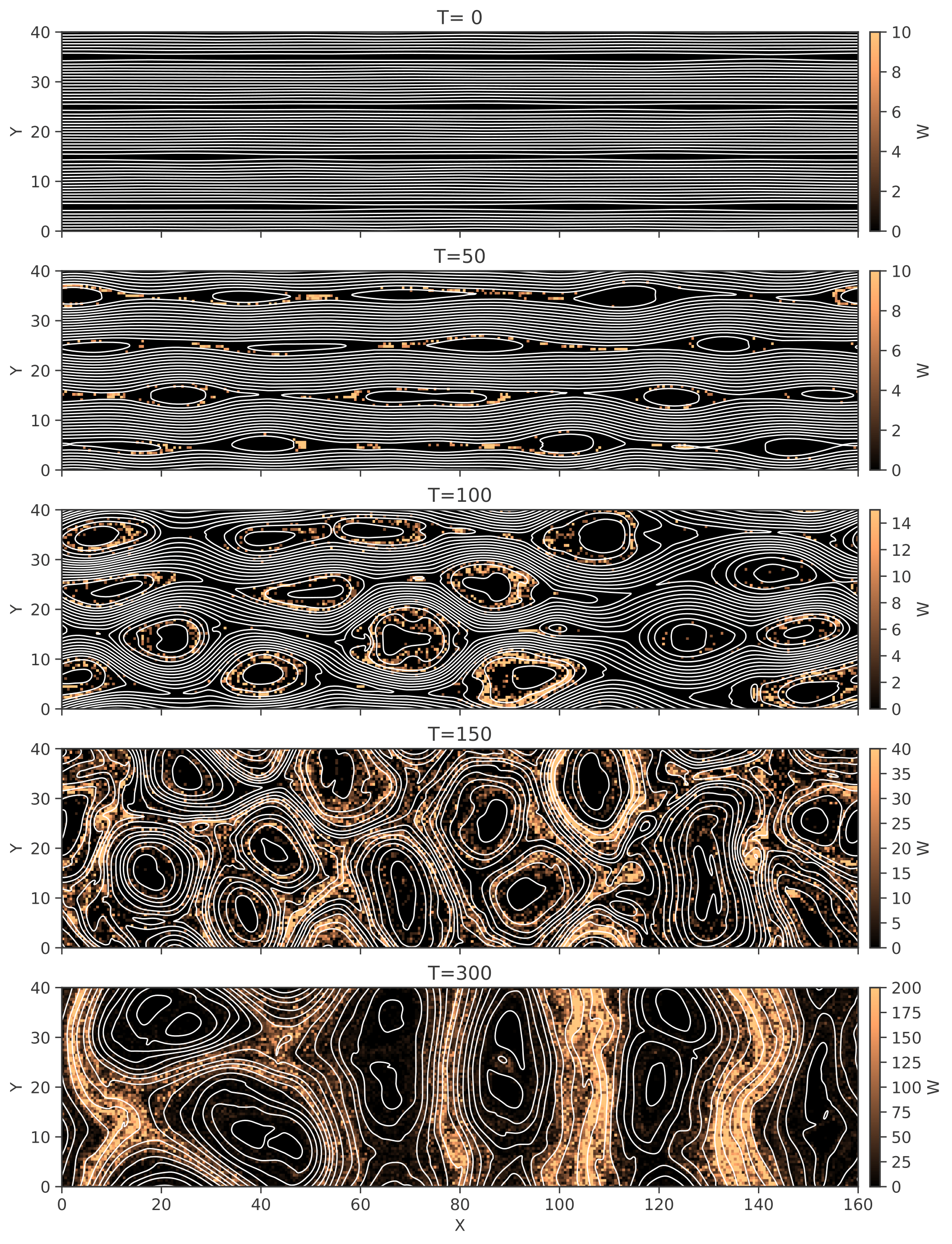}
		\end{center}
		\caption{Snapshots of the energy density of particles defined by $W=\int_{E_0}^\infty f(x,y,E)dE$ in the 2D case at different times, $t=0,50,100,150,300$. White lines corresponds to the magnetic field lines}
		\label{fig:2DWdist}
	\end{figure}

	The location of energetic particles depend on the stage of the evolution of multiple current sheets.
	Fig. \ref{fig:2DWdist} shows the time evolution of the energy density defined by,
	\begin{equation}
		W(x,y)=\int_{E_{min}}^\infty f(x,y,E)EdE,
	\end{equation}
	where $E_{min}$ is the minimum energy and set to $E_{min}=4$, so that we count only energetic particles.
	These panels correspond to different times, $t=0,50,100,150,300$ from top to bottom.
	The white lines are the magnetic fields lines.
	Note that the color scales are different at each time.
	It is obvious that there is no energetic particles at the initial time.
	After the onset of magnetic reconnection ($t=50$), there are some energetic particles produced along a current sheet.
	This acceleration is typical for magnetic reconnection \cite{oka2010,arnold2021}.
	As magnetic islands grow in size, we can see that energetic particles are trapped inside of magnetic islands. 
	At $t=150$, when magnetic islands interact with each other, it seems that energetic particles locate among magnetic islands rather than trapped inside.
	This is more evident at the end of the simulation ($t=300$), the energy density outside of magnetic islands is much higher than the inside.
	Therefore, we can conclude that energetic particles are firstly accelerated inside current sheets and trapped inside of magnetic islands, then released and further accelerated after magnetic islands start interacting with each other.
	Note that \citet{hoshino2012} also observed that energetic particles locate outside of magnetic islands in the simulation of multiple current sheets.
	
\begin{figure}[h!]
		\begin{center}
			\includegraphics[width=16cm]{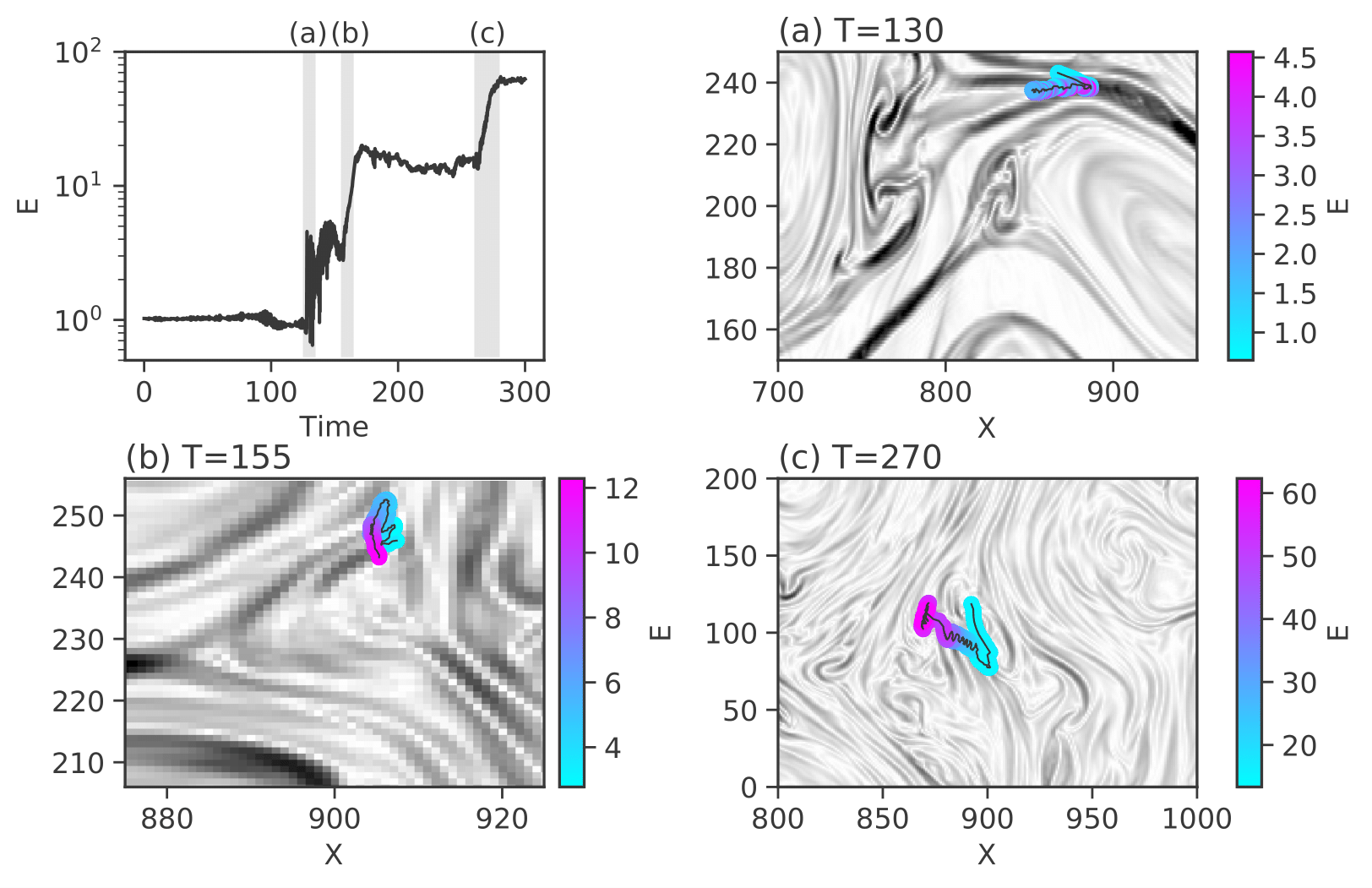}
		\end{center}
		\caption{A typical tranjectory of an accelerated particle in the 2D case. Top-left panel: time evolution of the particle energy, (a)-(c): particle trajectory (black line) and corresponding energy (color map) in the background of the current density $J_z$ (grey scale). Times (a)-(c) correspond to the shaded region in the top-left panel.}
		\label{fig:2Dptrj}
	\end{figure}

	Particles are efficiently accelerated by turbulence.
	In Fig. \ref{fig:2Dptrj}, the left-top panel shows the time evolution of the energy of a typically accelerated particle.
	The shaded regions denoted by (a)-(c) correspond to the other panels in Fig. \ref{fig:2Dptrj}
	The color scale in the panels (a)-(c) represents the particle energy.
	There are three major acceleration events, the first one is at $t=130$ and the acceleration is a quick energization.
	As seen in the panel (a), the particle is accelerated by a reconnection outflow of a single current sheet.
	When the particle enters a current sheet, it is kicked and moves along the outflow.
	The second ($t=155$) and third ($t=270$) accelerations are formally similar and accelerated by turbulence.
	As mentioned early, the turbulence is produced by the interaction of magnetic islands, and it starts from $T\sim150$.
	The motion of the particle appears stochastic in the panels (b) and (c), and the acceleration time is gradual compared to the first acceleration.
	The slopes of the two acceleration times are consistent.
	The particle energy finally reaches $E=60$.
	The particle trajectory indicates that, at first, a particle is energized in a single current sheet and then is further accelerated by turbulence produced due to the interaction of magnetic islands.
	
\begin{figure}[h!]
		\begin{center}
			\includegraphics[width=15cm]{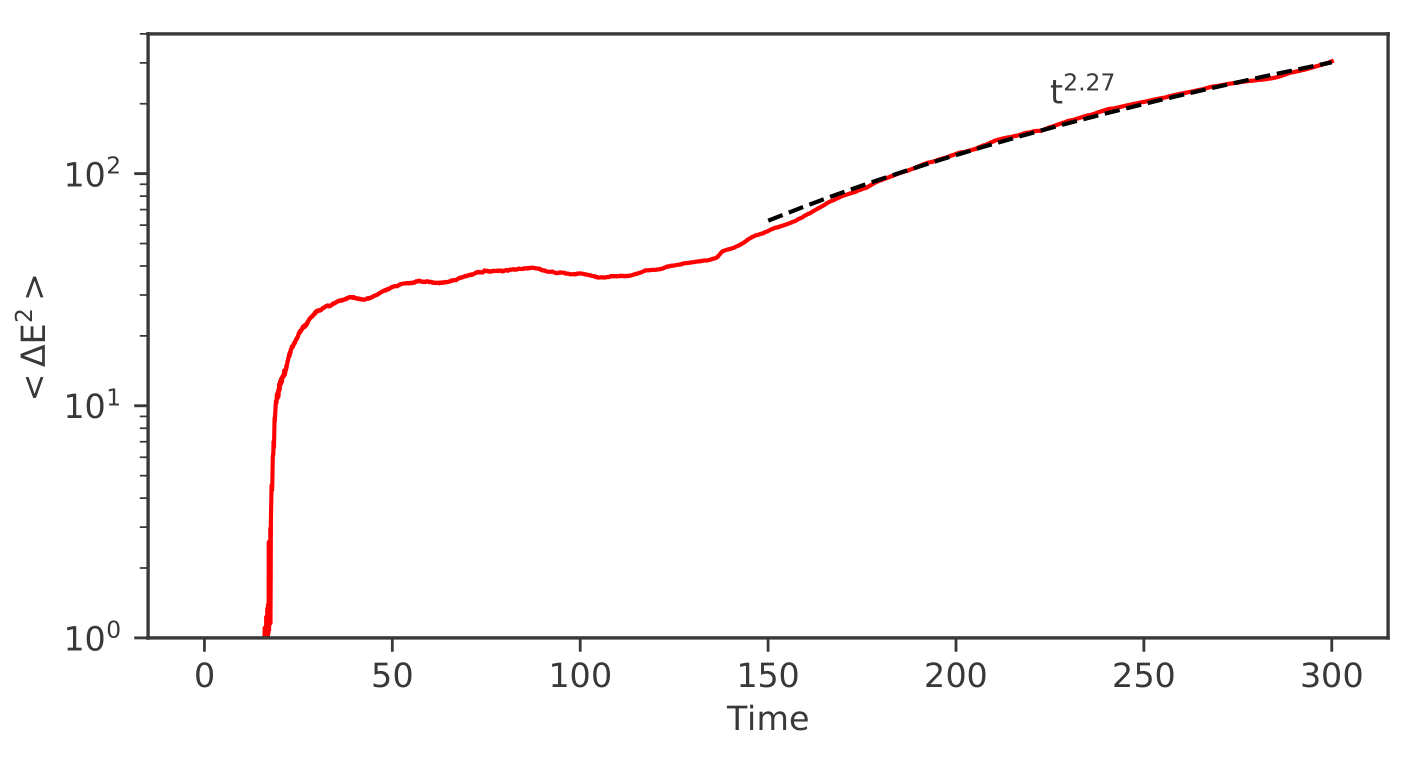}
		\end{center}
		\caption{Mean square displacement of the energy of energetic particles ($E>4$) of the 2D case. Balck dashed line is propotional to $t^{2.27}$.}
		\label{fig:2Dpdiff}
	\end{figure}

	The diffusion of energetic particles in the energy space is super-diffusive.
	Fig. \ref{fig:2Dpdiff} shows the mean square displacement (MSD) of  energy $<\Delta E^2>$ of energetic particles \cite{vlahos2008,sioulas2020}.
	We only consider particles whose energy is larger than $E=4$ since the motion of lower-energy particles may significantly change the MSD \cite{sioulas2020}.
	The definition of $<\Delta E^2>$ is as follows,
	\begin{gather}
		<\Delta E^2>=\frac{1}{N_p}\sum_{j=1}^{N_p}\left|\Delta E(t)\right|^2,
	\end{gather}
	where $N_p$ is the number of energetic particles ($E>4$).
	Here, $\Delta E(t)$ is the displacement of a particle energy,  $\Delta E(t)=E(t)-E(0)$ where $E(0)$ is the initial particle energy.
	However, since the number of energetc particles are a few until $t=20$ and $\sim10^4$ at $T\sim150$ (not shown here), we only consider the time after $t=150$.
	The MSD of energy can be fitted by a power-law $<\Delta E^2>\propto t^{a_E}$ with a power-law index of $a_E=2.27$.
	This indicates that the energy transport is super-diffusive.
	Note that the index $a_r<1$ is sub-diffusion and $a_r=1$ is normal diffusion.\\
	
	\subsection{3D case}
\begin{figure}[h!]
		\begin{center}
			\includegraphics[width=15cm]{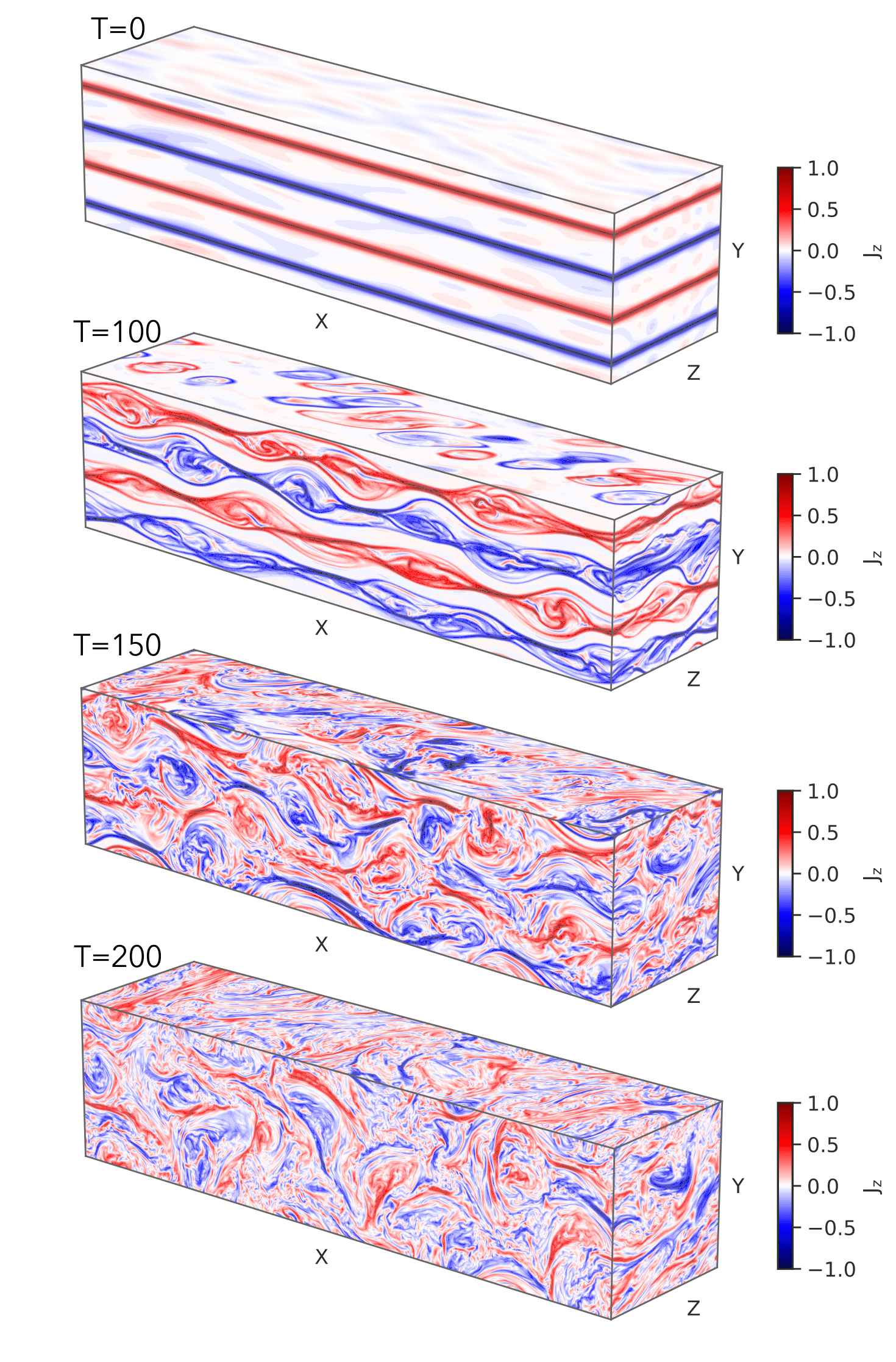}
		\end{center}
		\caption{Time evolution ($t=0,100,150,200$) of the current density $J_z$ in the 3D case.}
		\label{fig:3Dsnapshot}
	\end{figure}

	Multiple current sheets in the 3D simulation box become turbulent via the tearing-mode instability.
	The 3D simulation has the same condition as the 2D simulation but it extends the simulation box to the $z-$direction by $40L_0$ and we use a smaller value of the CFL number ($c_{CFL}=0.25$).
	Fig. \ref{fig:3Dsnapshot} shows the snapshots of the current density $J_z$ at the different times $t=0,100,150,200$.
	As the same in the 2D simulation, four current sheets are located inside of the simulation parallel to the $x-z$ plane at $t=0$.
	Small fluctuations seen at $t=0$ are because of the initial fluctuations defined by Eq. \ref{eq:3Dfluct}.
	The initial fluctuations initiate the tearing-mode instability, and magnetic reconnection proceeds at the current sheets.
	The location of the magnetic reconnection is also random as well as the 2D simulation.
	Although magnetic islands grow in size after the onset of magnetic reconnection, the shape of magnetic islands is not as clear as magnetic islands in the 2D simulation.
	This is because the magnetic reconnection occurs at random on the current sheets and magnetic islands merge with each other in the 3D simulation.
	Therefore, the evolution of current sheets is much more complicated in the 3D simulation.
	Due to the interaction of the destabilized current sheets, the system appears turbulent at $t=150$.
	At the end of the simulation ($t=200$), small scale fluctuations are more visible than at $t=150$, and the system transits to a highly-turbulent system.
	The turbulence appears isotropic since there is no background magnetic field.
	
\begin{figure}[h!]
		\begin{center}
			\includegraphics[width=15cm]{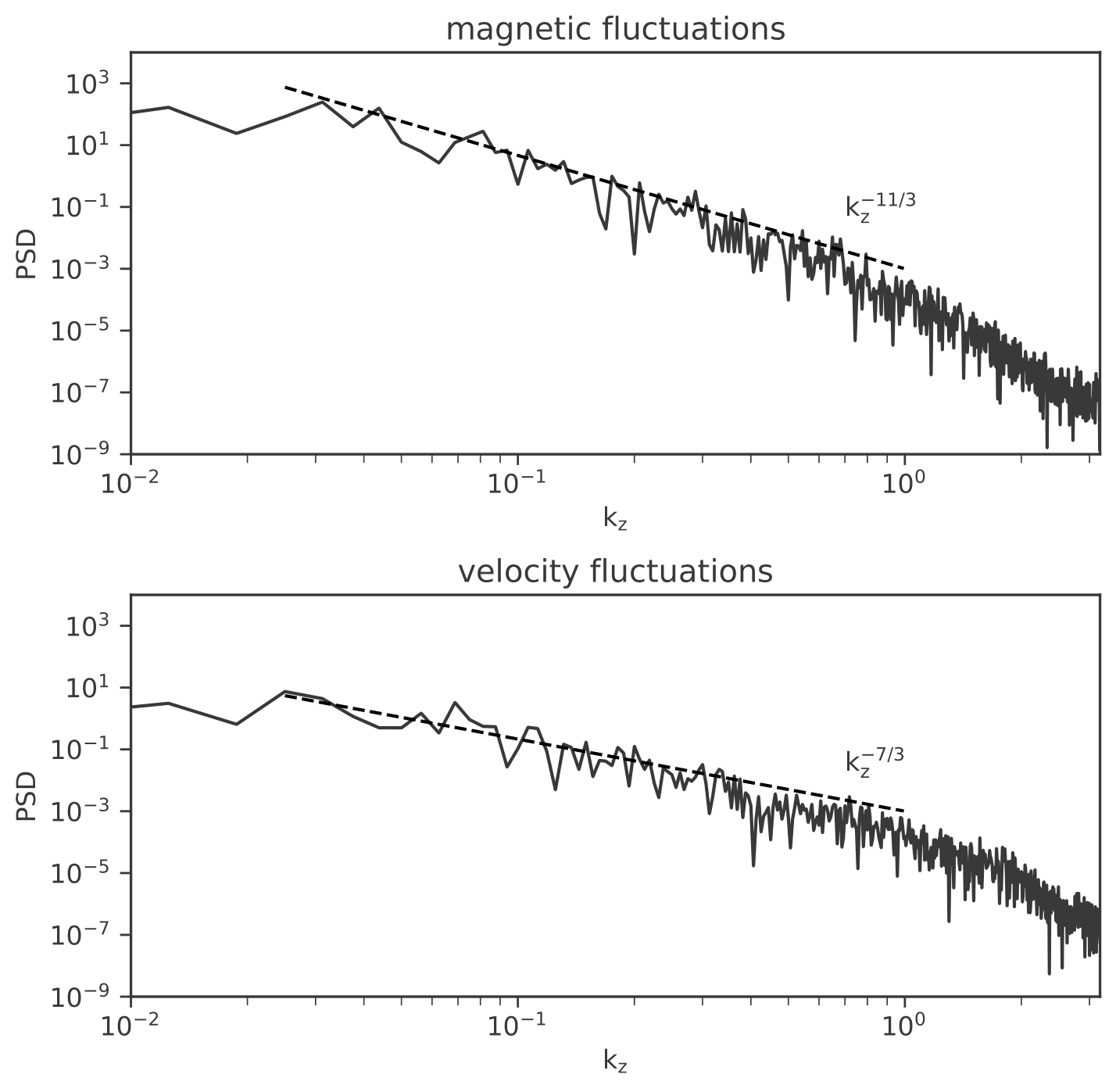}
		\end{center}
		\caption{Power spectrum density of magnetic and velocity fluctuations of the 3D case at $t=200$. }
		\label{fig:3DPSD}
	\end{figure}

	The spectra of the magnetic and velocity fluctuations form a $-11/3$ and $-7/3$ power-law, respectively.  
	Fig. \ref{fig:3DPSD} represents the PSD of magnetic (top panel) and velocity (bottom panel) fluctuations along the $x-$direction which is averaged over the $y-z$ plane.
	The magnetic PSD exhibits a $-11/3$ power-law over the range of $k_z\in[0.03,0.6]$, and the larger wavenumber range is dissipated by the artificial dissipation effects (resistivity and bulk viscosity). 
	On the other hand, the velocity PSD can be also fitted by a $-5/3$ power-law over the range $k_z\in[0.03, 0.7]$.
	The normalized cross helicity and residual energy are $8\times10^{-4}$ and $-0.53$, respectively.
	This indicates that the Alfv\'enic fluctuations are dominant over the velocity fluctuations.
	
\begin{figure}[h!]
		\begin{center}
			\includegraphics[width=15cm]{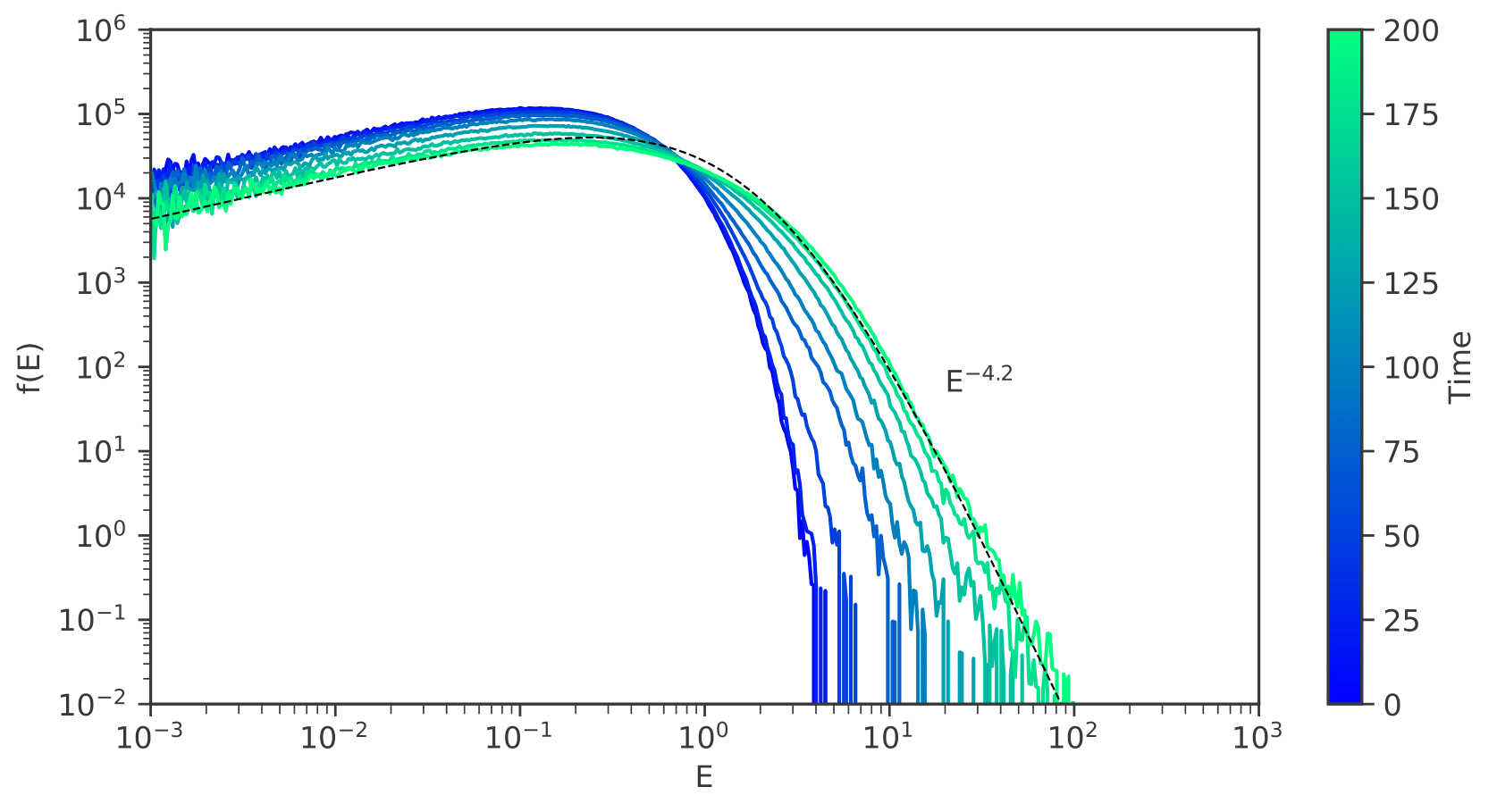}
		\end{center}
		\caption{Time evolution of the energy distribution of test particles in the 3D case. Black dashed line is a Kappa distribution with a temperature of $T_\kappa=0.8$ and a Kappa index of $\kappa=4.2$.}
		\label{fig:3Dedist}
	\end{figure}

	Non-thermal particles are produced during the evolution of the multiple current sheets and form a power-law tail.
	Fig. \ref{fig:3Dedist} shows the energy distribution of test particles at different times corresponding to the color scale.
	After the onset of the magnetic reconnection, the existence of non-thermal particles is not as obvious as in the 2D simulation.
	The particles seem to be heated rather than accelerated.
	However, a power-law tail starts forming after the turbulence begins to be created ($t\sim125$).
	At the end of the simulation ($t=200$), energetic particles appear forming a power-law tail with an index of $4.2$.
	The entire distribution is roughly fitted by a Kappa distribution (Eq. \ref{eq:kappa}) with a temperature of $T_\kappa=0.8$ and a Kappa index of $\kappa=4.2$.
	The maximum energy of accelerated particles is $\sim100E_0$.  
	
\begin{figure}[h!]
		\begin{center}
			\includegraphics[width=15cm]{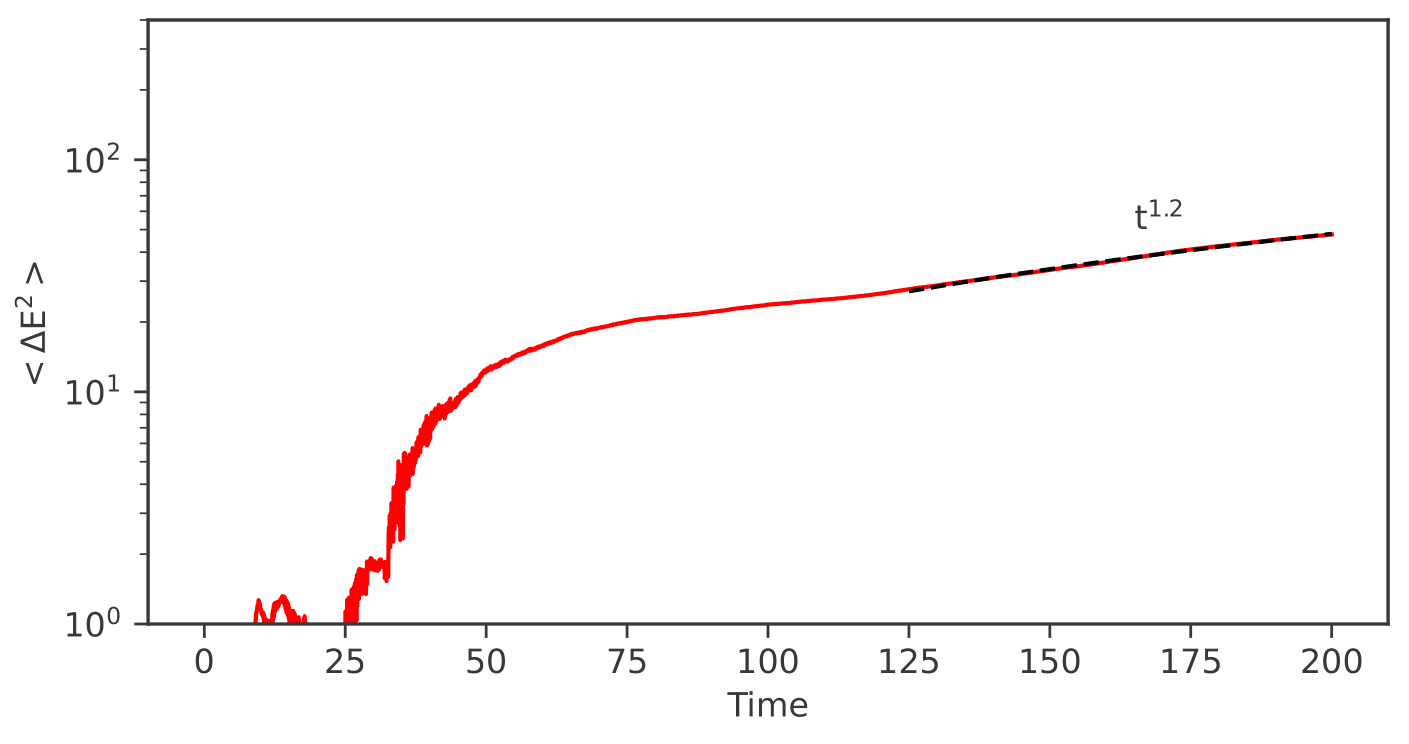}
		\end{center}
		\caption{Mean square displacement of the energy of energetic particles ($E>4$) of the 3D case. Balck dashed line is propotional to $t^{1.2}$.}
		\label{fig:3Dpdiff}
	\end{figure}

	Super-diffusion of energetic particles is observed in the energy space.
	Fig. \ref{fig:3Dpdiff} shows the time evolution of the energy MSD of energetic particles.
	We consider only particles whose energy is larger than $4$.
	The number of particles is a few until $t=50$.
	Therefore, we focus on the later time.
	After magnetic islands start interacting with each other ($t=125$), the MSD is fitted by $\propto t^{1.2}$.
	This indicates that the particle acceleration in the evolution of the turbulence is super-diffusive.

	\section{Discussion and Conclusion}
	Although the evolution of multiple current sheets is different in the 2D and 3D simulations, both cases yield turbulence at the end of the simulation.
	In the 2D case, current sheets are unstable for the tearing-mode instabiliy, and magnetic islands are produced by magnetic reconnection.
	In the 3D case, magnetic reconnection occurs at random on current sheets (the $x-z$ plane) and the evolution of magnetic islands differs along the $z-$direction. 
	This makes the evolution of current sheets more complicated in the 3D case than in the 2D case.
	However, the system of both cases develops into a highly turbulent state at the end of the simulation.
	
	The efficiency of particle acceleration in the 2D simulation is higher than that in the 3D simulation.
	While the power-law index of the energy distribution in the 2D case exhibits $-2.2$, it is $-4.2$ in the 3D case.
	This simply implies that the acceleration in the 2D case is more efficient than in the 3D case.
	The maximum energy of the 2D case ($E_{max}\sim300E_0$) is also higher than that of the 3D case ($E_{max}\sim100E_0$).
	We note that the power-law tails extend to the maximum energies.
	The index of the observed super-diffusion in the 2D case ($2.27$) is higher than that in the 3D case ($1.2$).
	This also indicates that the 2D acceleration is more efficient than the 3D acceleration.
	We interpret this because in the 2D case particles can be easily trapped in the turbulence more than in the 3D case.
	
	Compared to previous studies, an MHD scale system of multiple current sheets is an efficient acceleration site.
	Former kinetic simulations \cite{drake2010,burgess2016,nakanotani2021} did not show a significant particle acceleration, for example, (1) no power-law tail and (2) acceleration by a factor of a few decades.
	However, as we have shown, the particle acceleration in the both 2D and 3D cases forms a power-law tail and is accelerated by a factor of more than $100$.
	Therefore, we conclude that particle acceleration in an MHD-scale system of multiple current sheets is efficient.
	Although it is possible to directly verify this by extending kinetic simulations to MHD-scale, it may not be realistic due to the current computational power.
	
	The particle acceleration observed in the 2D and 3D cases can be modeled by a fractional Fokker-Planck model, which is a generalization of a classical Fokker-Planck model.
	It is thought a super-diffusion in the energy space is an indication of efficient particle acceleration and can be related to the formation of a power-law tail \cite{vlahos2004, isliker2017, isliker2019, sioulas2020}.
	There are several models for anomalous diffusions in the energy space as well as the real space using a fractional Fokker-Planck model to understand particle acceleration with the nature of anomalous diffusion often observed in space plasmas \cite{milovanov2001, vlahos2004, bian2008, isliker2017, leroux2021}.
	In a future study, we use a fractional Fokker-Planck model and compare it with several simulations by varying the background magnetic field.
	
	We do not expect a plasma beta dependence on particle acceleration.
	Following the fact that plasma beta does not strongly affect the tearing-mode instability \cite{landi2008}, We assume that multiple current sheets develop into a turbulent state in several values of plasma beta.
	Since the structure of magnetic reconnection appears to be turbulent in a low-beta plasma \cite{zenitani2015,zenitani2020}, we anticipate that particles can be still efficiently accelerated by the turbulence as well as we have seen in our simulations. 
	
	On the other hand, we expect some dependencies of particle acceleration on the strength of the background magnetic field.
	Several studies of magnetic reconnection show that particle acceleration becomes less efficient as the background magnetic field becomes strong \cite{fu2006,huang2016,werner2017,arnold2021}.
	This can be because particle motion for Fermi acceleration is limited by the background magnetic field.
	In the system of multiple current sheets with a strong magnetic field, an initial acceleration by a single magnetic reconnection becomes less efficient, and therefore the latter acceleration phase due to turbulence can be less efficient as well. 
	
	In conclusion, We have performed 2D and 3D MHD simulations of multiple current sheets combined with test particle simulations to investigate particle acceleration.
	In the both cases, multiple current sheets are unstable for the tearing-mode instability and transit to a turbulence state with a power-law spectra for magnetic and velocity fluctuations.
	We also observe the formulation of magnetic islands because of magnetic reconnection during the transition.
	Non-thermal particles are efficiently produced due to turbulence generated by the interaction of magnetic islands. 
	Their energy distribution can be fitted by a Kappa distribution with a Kappa index (or power-law index) of $-2.2$ and $-4.2$ for the 2D and 3D case, respectively.
	The efficient acceleration is consistent with the observed super-diffusion in the energy space for the both cases, which can be modeled by a fractional Fokker-Planck model.

	\section*{Conflict of Interest Statement}
	The authors declare that the research was conducted in the absence of any commercial or financial relationships that could be construed as a potential conflict of interest.
	
	\section*{Author Contributions}
	The Author Contributions section is mandatory for all articles, including articles by sole authors. If an appropriate statement is not provided on submission, a standard one will be inserted during the production process. The Author Contributions statement must describe the contributions of individual authors referred to by their initials and, in doing so, all authors agree to be accountable for the content of the work. Please see  \href{http://home.frontiersin.org/about/author-guidelines#AuthorandContributors}{here} for full authorship criteria.
	
	\section*{Funding}
	Details of all funding sources should be provided, including grant numbers if applicable. Please ensure to add all necessary funding information, as after publication this is no longer possible.
	
	\section*{Acknowledgments}
	This is a short text to acknowledge the contributions of specific colleagues, institutions, or agencies that aided the efforts of the authors.
	
	
	\section*{Data Availability Statement}
	The simulation data shown in this paper are made available upon request by the corresponding author.
	
	\bibliographystyle{frontiersinHLTH&FPHY} 
	\bibliography{reference}
	
	
	\clearpage
	\section*{Figure captions}
	
\end{document}